\documentclass[lettersize,journal]{IEEEtran}
\usepackage{etoolbox}
\makeatletter
\patchcmd{\@makecaption}
{\scshape}
{}
{}
{}
\makeatletter
\patchcmd{\@makecaption}
{\\}
{.\ }
{}
{}
\makeatother
\usepackage{amsmath,amssymb}
\usepackage{subcaption}
\usepackage{graphicx,graphics,color,psfrag}
\usepackage{cite,balance}
\usepackage{caption}
\allowdisplaybreaks
\usepackage{accents}
\usepackage{amsthm}
\usepackage{bm}
\usepackage[english]{babel}
\usepackage{multirow}
\usepackage{enumerate}
\usepackage{cases}
\usepackage{stfloats}
\usepackage{dsfont}
\usepackage{color,soul}
\usepackage{amsfonts}
\usepackage{cite,graphicx,amsmath,amssymb}
\usepackage{fancyhdr}
\usepackage{hhline}
\usepackage{graphicx,graphics}
\usepackage{array,color}
\usepackage{amsmath}
\usepackage{float}
\usepackage{amssymb}
\usepackage{amsmath}
\usepackage{amsthm}
\usepackage{amsfonts}
\usepackage{graphicx}
\usepackage[ruled]{algorithm2e}
\usepackage{epstopdf}
\usepackage{cite}
\usepackage{amsmath,bm}
\usepackage{graphicx}
\usepackage{color}
\usepackage{graphicx}
\usepackage{calc}
\usepackage{caption}
\usepackage{diagbox}

\usepackage{verbatim}

\usepackage{booktabs}

\usepackage{algpseudocode}
\usepackage{amsmath}
\usepackage[numbers,sort&compress]{natbib}

\begin{document}

\title{Autoencoder-based Semantic Communication Systems with Relay Channels}

\author{Xinlai Luo, Zhiyong Chen, Bin Xia, and Jiangzhou Wang~\IEEEmembership{Fellow, IEEE}
        % <-this % stops a space
\thanks{X. Luo, Z. Chen, and B. Xia are with the School of Electronic Information and Electrical Engineering, Shanghai Jiao Tong University, China (e-mail: \{newcomer, zhiyongchen, bxia\}@sjtu.edu.cn). J. Wang is with the School of Engineering, University of Kent, Canterbury CT2 7NT, U.K. (e-mail: j.z.wang@kent.ac.uk).}% <-this % stops a space
%\thanks{Manuscript received April 19, 2021; revised August 16, 2021.}
}

{%Shell \MakeLowercase{\textit{et al.}}: A Sample Article Using IEEEtran.cls for IEEE Journals}


\maketitle

\begin{abstract}
In this letter, we propose a semantic communication scheme for wireless relay channels based on Autoencoder, named AESC, which encodes and decodes sentences from the semantic dimension. The Autoencoder module provides anti-noise performance for the system. Meanwhile, a novel semantic forward (SF) mode is designed for the relay node to forward the semantic information at the semantic level, especially for the scenarios that there is no common knowledge shared between the source and destination nodes. Numerical results show that the AESC achieves better stability performance than the traditional communication schemes, and the proposed SF mode provides a significant performance gain compared to the traditional forward protocols.
\end{abstract}

\begin{IEEEkeywords}
Semantic Communication, Autoencoder, Relay, Semantic Forward.
\end{IEEEkeywords}

\section{Introduction}
\IEEEPARstart{S}{emantic} communication was considered on the second level of communication problem \cite{1949The}, aiming to convey the semantic information of the transmission symbols accurately, instead of accurately recovering the transmitted information. With the development of artificial intelligence (AI) technology, semantic communication has attracted one's attention again and is considered as one of the future key mobile technologies.

Recently, several semantic communication concepts have been proposed, i.e., semantic-oriented and goal-oriented communication \cite{6g}, semantic-aware networking based on federated edge intelligence \cite{2020semantic}. Besides reconstructing a new paradigm of semantic communication from a mathematical perspective \cite{towards}, many innovative communication schemes have been developed based on neural networks (NNs) and semantic interpretation modules to replace conventional communication blocks. The work \cite{air} developed a point-to-point communications system whose entire physical layer processing was performed by NNs, but it is difficult to overcome the difference between the actual channel and the channel model used for training. In \cite{gan}, a conditional generative adversarial net (GAN) was designed to represent channel effects. Moreover, based on a semantic autoencoder (SAE) \cite{auto}, the network can learn a projection function from feature space to a semantic embedding space in zero-shot learning (ZSL) models. Based on Transformer \cite{attention} which is the dominating language model in natural language processing (NLP), authors developed a deep learning based semantic communication system, named DeepSC \cite{semantic}. It has better performance in the low signal-to-noise ratio (SNR) regime compared with the traditional communication system. However, the commonly used cross-entropy (CE) loss was flawed. In \cite{2021rethinking}, reinforcement learning (RL) was used to narrow the semantic distance and an RL-based similarity-targeted semantic communication mechanism was established. Furthermore, it is worth noting that most of the existing semantic communication only considered end-to-end (E2E) communications system, and did not consider cooperative communications, e.g., wireless relay channel. Cooperative communication systems have been widely used in the wireless communication system \cite{coopera}, so it is important to study semantic communication for the cooperative communication system.

In this paper, we try to design a novel Autoencoder-based semantic communication system (AESC) to improve the reliability of semantic communication between multiple nodes. Firstly, we introduce a reliable autoencoder in the semantic communications system, including encoder and decoder powered by NNs, to compress semantic information and resist channel noise interference. Secondly, although the semantic communication is based on the common semantic background knowledge (BK) of the transmitter and receiver, in practice, it is difficult for the source node and the destination node to have exactly the same BK. As is well known, the traditional relay node can help establish the communication between a source and a destination. Thus, we propose a new semantic forward (SF) protocol in the relay node to solve the problem. With the SF protocol, the relay node can obtain semantic information from the source node based on the BK between the relay node and the source node, and then forward the semantic information based on the BK between the relay node and the destination node, thereby helping the source node and the destination node to  communicate accurately at the semantic level under different background knowledge.

 %The traditional relay forward modes, the amplify-and-forward (AF) mode and the decode-and-forward (DF) mode, cannot perform cooperative communication at the semantic level.

\begin{figure*}[t]
	\begin{center}
		\includegraphics[width=18cm]{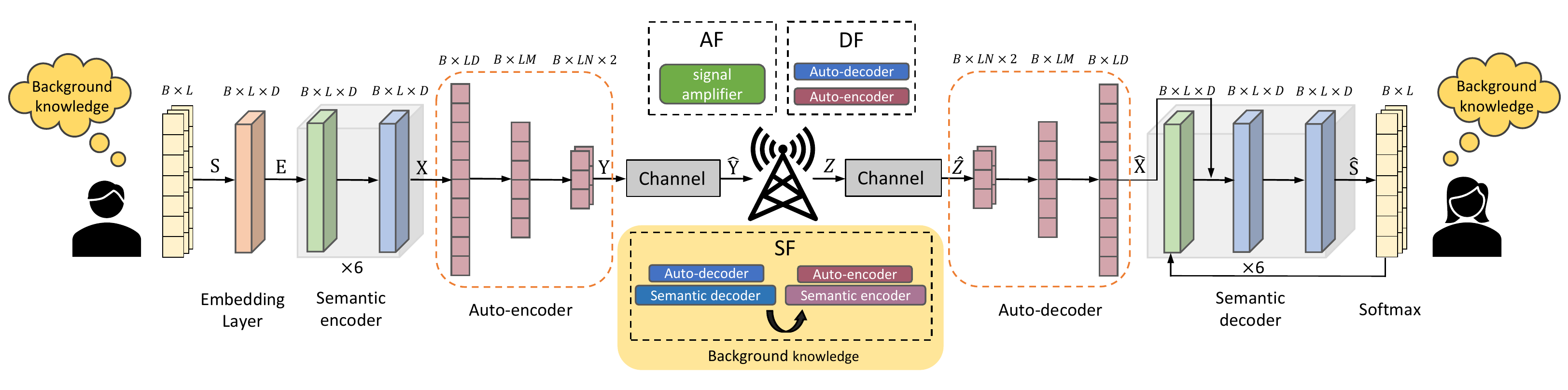}
	\end{center}
	\caption{\small{A wireless relay semantic communication model.}
	}
	\label{model1}
\end{figure*}

\section{System Model}
Fig. \ref{model1} shows an Autoencoder-based semantic communication with one-way relay channels, where the source node transmits the information to the destination node through a relay node based on the semantic communication. The semantic communication consists of two levels: semantic level and transmission level. The semantic level contains the semantic coding layer and the semantic decoding layer for extracting and analyzing the semantic information, respectively. The transmission level guarantees that semantic information can be transmitted accurately in the wireless channel.
\subsection{Autoencoder-based Semantic Communications}
\subsubsection{Encode Layer}
The input of the AESC in the source node is a sentence $\mathbf{s} = [w_1, w_2, \cdots, w_L]$, $\mathbf{s} \in \mathcal{K}$, where $w_l$ represents the $l$-th word in the sentence and $\mathcal{K}$ is the background knowledge. As shown in Fig. \ref{model1}, the input sentence $\mathbf{s}$ is embedded into a vector $\mathbf{e}$, and each word $w$ in the sentence is mapped to $D$ dimension. Then, we apply semantic encoder which is the encoder of Transformer to encode $\mathbf{e}$ to a semantic vector $\mathbf{x}$. The Transformer encoder consists of two parts, named self-attention sublayer and feed-forward sublayer.
%The self-attention sublayer can capture the implicit semantic dependency between words in the sentence, and in order to capture the semantic features of the sentence as well as possible, we use multi-head attention the same as Transformer in NLP. The feed-forward sublayer is a fully connected layer independently applied to words in each position of an input sentence.%
Connecting the above two parts alternately six times forms a complete semantic encoder network.

Then the source node passes the semantically encoded vector $\mathbf{x}$ into the Auto-encoder to get compressed vector $\mathbf{y}$. The Auto-encoder has functions similar to traditional channel coding, has anti-noise and encryption capabilities, and can compress the semantic vector to a certain extent.

Accordingly, the receive signal at the relay node is
\begin{equation}
	\label{eq:C}
	\widehat{\mathbf{y}} = h_1\mathbf{y} + \mathbf{n}_1,
\end{equation}
where $h_{1}$ represents the Rayleigh fading channel between the source node and the relay node, and $\mathbf{n_1}\sim\mathcal{C}\mathcal{N}(0, \sigma^{2}_{n})$ is the additive white Gaussian noise (AWGN).
%\subsubsection{Relay Node}

Next, the relay node can forward the information to the destination node based on the forward protocol, which will be discussed in the next subsection.
% For the AF scheme, the signal received at the relay node is directly amplified and forwarded to the destination node.
%%We thus have $z=\alpha \widehat{\mathbf{y}}$ for the AF scheme, where $\alpha$ is an amplification constant factor which is chosen to satisfy the total power constraint at the relay node.
%For the DF scheme, the relay node can decode and forward the received information at different layers based on its BK.
\subsubsection{Decode Layer}
As shown in Fig. \ref{model1}, the signal received at the destination node is given by
\begin{equation}
	\label{eq:C}
	\widehat{\mathbf{z}} = h_2\mathbf{z} + \mathbf{n}_2,
\end{equation}
where $h_{2}$ represents the Rayleigh fading channel between the
relay node and the destination node, and $\mathbf{n_2}\sim\mathcal{C}\mathcal{N}(0, \sigma^{2}_{n})$ is AWGN.

The destination node gets $\widehat{\mathbf{z}}$ and passes it through the Auto-decoder to recover the semantic vector $\widehat{\mathbf{x}}$ which keeps the semantic information of $\mathbf{x}$. Finally, we input $\widehat{\mathbf{x}}$ into the semantic decoder to recover the transmitted sentences. The semantic decoder is divided into three sublayers, one more layer than the Transformer encoder. Between self-attention sublayer and feed-forward sublayer is an encoder-decoder attention sublayer that can help the decoder focus on the relevant parts of the input sentence. In addition, the semantic decoder has two inputs. One is the semantic vector recovered by the softmax layer and another is the transmitted sentence according to the framework of Transformer.

%where $\left|h_{1}\right|\left(\left|h_{2}\right|\right)$ is Rayleigh distributed and $\left|h_{1}\right|^{2}(\left|h_{2}\right|^{2})$ can be easily proved to be exponential random variable with parameter $\lambda_{1}\left(\lambda_{2}\right)$. $\mathbf{n_1}\left(\mathbf{n_2}\right)\sim\mathcal{C}\mathcal{N}(0, \sigma^{2}_{n})$ is the additive white Gaussian noise (AWGN).

%However, different from the translation task in NLP that we can get the complete information of the input sentence on the decoding side, the destination node has no idea what sentence was sent from the source node. Thus, we set the initial sentence input as a vector of $0$. That is, the information of the transmitted sentence is none, and then only the semantic vector recovered by the Auto-decoder is used for decoding. The word with the highest probability in the first position can be extracted after the softmax layer, and then the acquired information is used as a new input vector for the second time decoding, obtaining the second word of the sentence. Then continue with the previous operation and gradually shift to the right by one position to recover the complete transmitted sentence. We call this decoding network as Transformer Greedy-decoder.%

\subsection{Relay Forward Protocol}
The forward protocol of the relay node can be divided into two scenarios according to whether the relay node has the semantic BK.
\subsubsection{Traditional relay mode}
The relay node only processes the signal at the physical level and does not involve semantics. If the amplify-and-forward (AF) mode is used, the signal received at the relay node is directly amplified and forwarded to the destination node, and we thus have $z=\alpha \widehat{\mathbf{y}}$. Here, $\alpha$ is an amplification constant factor that is chosen to satisfy the total power constraint at the relay node. If the relay node uses the decode-and-forward (DF) mode, the relay node can decode and forward the received information at different layers based on its BK. If the relay node does not have the BK shared by the source node and the destination node, it can not further perform semantic decoding; otherwise, the relay node can perform semantic decoding.
\subsubsection{Semantic relay mode}
The traditional relay mode requires the source node and destination node to have the same BK. However, it is difficult for the source node and the destination node to have exactly the same BK. To overcome this problem, we design a novel relay forward mode here, named semantic forward. The key idea of SF is that the relay node can perform the semantic decoding to obtain  the semantic information $\mathbf{s}$ from the received signal $\widehat{\mathbf{y}} $ based on the BK between the source node and itself, and then it can recode the sentences $\mathbf{s}$  in a way that the destination node can understand based on another BK between the source node and the relay node. For example, the source node said:`` My son is very good at CS'', but the BK of the destination defaults that CS is a game called Counter-Strike and the destination node knows the name of the source node's son is Bob. In this case, the advantage of the proposed SF can be reflected at this time. The relay node can know that CS said by the source node means computer science and the destination node knows Bob. Based on the SF protocol, the destination node receives the sentence:``Bob is very good at computer science''. We can see from this example that SF not only reduces semantic distortion but also performs semantic compression.
\section{Model Training}
Different from Fig. \ref{model1}, we can train each module in the E2E model. At this time, $\widehat{\mathbf{z}} = \widehat{\mathbf{y}}$, the input of the whole network is a sentence $\mathbf{s}$. The encoded semantic vector can be represented by
\begin{equation}
\mathbf{y} = F_{\beta}(S_{\alpha}(\mathbf{s})),
\end{equation}
where $S_{\alpha}(\mathbf{\cdot})$ is the semantic encoder network with the parameter set $\alpha$ and $F_{\beta}(\mathbf{\cdot})$ is the Auto-encoder network with the parameter set $\beta$. The decoded sentence can be represented as
\begin{equation}
\widehat{\mathbf{s}} = G_{\eta}(D_{\theta}(\widehat{\mathbf{y}})),
\end{equation}
where $D_{\theta}(\mathbf{\cdot})$ is the Auto-decoder network with the parameter set $\theta$ and $G_{\eta}(\mathbf{\cdot})$ is the semantic decoder network with the parameter set $\eta$.
\subsection{Training Auto-encoder/decoder}
Different from the joint training of source and channel coding, we train the Autoencoder module separately to deal with different channel conditions more flexibly. The training of Auto-encoder and Auto-decoder does not require a true semantic vector as input. The function of this part is to make the distortion of the transmitted vector as small as possible. Therefore, we use randomly generated vectors for training, and the range of random vectors is $\left[ -2, 2 \right]$, similar to the encoded semantic vector. The Auto-encoder has two hidden layers, each layer compresses the semantic vector to a certain extent. In this paper, the first layer is compressed to {M} dimensions, and the second layer is compressed to {2N} dimensions, mapping into the {N} symbols of the real part plus the imaginary part. Obviously, the Auto-decoder has a symmetrical structure. Because the input and output are two vectors, we use the mean square error (MSE) as the loss function to reduce the distortion. Based on the transmitted vector $\mathbf{x}$ and the distorted vector $\widehat{\mathbf{x}}$, the loss function can be expressed as
\begin{equation}
	\label{mse}
	\mathcal{L}_{MSE} = \frac{\sum_{i=1}^{n}\left(x_{i}-\widehat{x}_{i}\right)^{2}}{N},
\end{equation}
where $x_{i}$ and $\widehat{x}_{i}$ represent the components of vector $\mathbf{x}$ and $\widehat{\mathbf{x}}$ respectively, $N$ is vector size. Finally, the ADAM algorithm is exploited to optimize the parameters $\beta, \theta$. The training algorithm is outlined in Algorithm \ref{channel}.

\begin{algorithm}[t]
	\caption{ Train Auto-encoder/decoder network.}
	\label{channel}
	\begin{algorithmic}[1]
		\Require
		Channel SNR value and a set of randomly generated vectors $X$;
		\Ensure
		Network $F_{\beta}(\mathbf{\cdot})$, $D_{\theta}(\mathbf{\cdot})$;
		\State Take a batch $\mathbf{x}$ from the set $X$;
		\label{code:1}
		\State $F_{\beta}(\mathbf{x}) \to \mathbf{y}$;
		\label{code:2}
		\State Transmit $\mathbf{y}$ over the Rayleigh fading channel: $\widehat{\mathbf{y}} = h\mathbf{y} + \mathbf{n}$;
		\label{code:3}
		\State $D_{\theta}(\widehat{\mathbf{y}}) \to \widehat{\mathbf{x}}$;
		\label{code:4}
		\State Compute loss $\mathcal{L}_{MSE}$ by (\ref{mse});
		\label{code:5}
		\State Gradient descent update $\beta, \theta$;
		\label{code:6}\\
		\Return $F_{\beta}(\mathbf{\cdot})$, $D_{\theta}(\mathbf{\cdot})$.
	\end{algorithmic}
\end{algorithm}
\begin{algorithm}[htb]
	\caption{ Train semantic encoder/decoder network.}
	\label{whole}
	\begin{algorithmic}[1]
		\Require
		Channel SNR value and the backgroud knowledge set $\mathcal{K}$ ;
		\Ensure
		Network $S_{\alpha}(\mathbf{\cdot})$, $G_{\eta}(\mathbf{\cdot})$;
		\State Load the pre-trained network $F_{\beta}(\mathbf{\cdot})$, $D_{\theta}(\mathbf{\cdot})$;
		\label{code:1}
		\State Take a batch $\mathbf{s}$ from the set $\mathcal{K}$ and embed $\mathbf{s}$;
		\label{code:2}
		\State $S_{\alpha}(\mathbf{s}) \to \mathbf{x}$;
		\label{code:3}
		\State $F_{\beta}(\mathbf{x}) \to \mathbf{y}$;
		\label{code:4}
		\State Transmit $\mathbf{y}$ over the Rayleigh fading channel: $\widehat{\mathbf{y}} = h\mathbf{y} + \mathbf{n}$;
		\label{code:5}
		\State $D_{\theta}(\widehat{\mathbf{y}}) \to \widehat{\mathbf{x}}$;
		\label{code:6}
		\State $G_{\eta}(\widehat{\mathbf{x}}) \to \widehat{\mathbf{s}}$;
		\label{code:7}
		\State Compute loss $\mathcal{L}_{CE}$ by (\ref{ce});
		\label{code:8}
		\State Gradient descent update $\alpha, \eta$;
		\label{code:9}\\
		\Return  $S_{\alpha}(\mathbf{\cdot})$, $G_{\eta}(\mathbf{\cdot})$.
	\end{algorithmic}
\end{algorithm}
\subsection{Training Semantic Encoder/Decoder}
The semantic encoder and decoder training process are illustrated in Algorithm \ref{whole}. Firstly, we load the pre-trained network shown in Algorithm \ref{channel}, and fetch minibatch $B$ from BK $\mathcal{K}$ to generate sentences $\mathrm{S} \in \Re^{B \times L}$, where $B$ is the batch size and $L$ is the length of sentences. Through the embedding layer, the sentences can be represented as a vector $\mathrm{E} \in \Re^{B \times L \times D}$, where $D$ is the embedding dimension of each word. After the transmission of the entire communication model, the reconstructed sentence $\widehat{\mathrm{S}} \in \Re^{B \times L}$ will be obtained. The cross-entropy is used as the loss function to measure the difference between $\mathrm{S}$ and $\widehat{\mathrm{S}}$, which can be formulated as
\begin{equation}
 \label{ce}
 \mathcal{L}_{CE}=-\frac{1}{m} \sum_{i=1}^{m} \sum_{j=1}^{l} p\left(x_{i j}\right) \log \left(q\left(x_{i j}\right)\right),
\end{equation}
where $p\left(x_{i j}\right)$ is the real probability that the $j$-th word in $i$-th sample, and $q\left(x_{i j}\right)$ is the predicted probability that the $i$-th word in $i$-th sample, and $m$ represents the number of samples in one batch. The $\mathrm{CE}$ can measure the difference between two probability distributions. By reducing the loss value of $\mathrm{CE}$ and optimizing parameters $\alpha, \eta$ by the stochastic gradient descent (SGD), the whole network can be robust to distortion.

\section{Performance Evaluation}
\subsection{Performance Metrics}
%In the traditional wireless communication system, bit-error-rate (BER) is usually considered as the optimization goal. For semantic communication, BER cannot reflect the performance of communication well, because, in many semantic communication scenarios, such as text chat, goal-oriented communications, we do not care whether the transmitted symbols are completely correct. What we need is the meaning behinds the bits.

For the semantic communication, the symbol has no value unless it has exploitable semantic information. Thus, we use two evaluation indicators to judge communication quality in this letter. One is the bilingual evaluation understudy (BLEU) score \cite{bleu} and the other one is semantic vector similarity.
\subsubsection{BLEU Score}
BLEU adopts the $k$-gram precisions rule, which calculates the similarity of $k$ groups of words between the candidate sentence and the reference sentence. For example, we use ``It is a nice day today" as the candidate sentence, and the reference sentence is ``Today is a nice day". If we use 1-gram precision, we can see that the candidate has a total of 6 words, and 5 words all hit the reference, then its 1-gram precision is 5/6. Its 3-gram precision is 2/4.

Firstly, we compute the brevity penalty BP as following
\begin{equation}
	\mathrm{BP}=\left\{\begin{array}{ll}
		1 & \text { if } c>r \\
		e^{(1-r / c)} & \text { if } c \leq r
	\end{array} ,\right.
\end{equation}
where $c$ is the length of the candidate translation and $r$ is the effective reference corpus length.
%Firstly, we compute the brevity penalty BP if the length of the candidate translation
%is shorter than the effective reference corpus length.
Secondly, we compute the geometric average of the modified $k$ -gram precisions $p_{k}$ by using positive weights $w_{k}$. Then, we have
\begin{equation}
\mathrm{BLEU}=\mathrm{BP} \cdot \exp \left(\sum_{k=1}^{K} w_{k} \log p_{k}\right).
\end{equation}
%Take $\log$ to make the ranking more obvious,
%\begin{equation}
%\log \mathrm{BLEU}=\min \left(1-\frac{r}{c}, 0\right)+\sum_{n=1}^{N} w_{n} \log p_{n}
%\end{equation}
The BLEU metric ranges from 0 to 1. It will attain a score of 1 if candidate sentences are identical to reference sentences. In this paper, we use the 2-grams BLEU score.
%\begin{figure}[t]
%	\begin{center}
%		\includegraphics[width=8cm]{modelaf.pdf}
%	\end{center}
%	\caption{\small{A wireless AF relay semantic communication model.}
%	}
%	\label{modelaf}
%\end{figure}
\subsubsection{Semantic Vector Similarity} In the task of NLP, the cosine similarity formula is often used to calculate the similarity of two-word vectors. Cosine similarity uses the cosine value of the angle between two vectors in the vector space as a measure of the difference between two individuals. The closer the cosine value is to 1, the closer the angle is to 0 degrees, the more similar the two vectors are.

Given two vectors, $A$ and $B$, the cosine similarity can be given by
\begin{equation}
\text { similarity }=\frac{\sum_{i=1}^{n} A_{i} \times B_{i}}{\sqrt{\sum_{i=1}^{n}\left(A_{i}\right)^{2}} \times \sqrt{\sum_{i=1}^{n}\left(B_{i}\right)^{2}}}.
\end{equation}
Here $A_{i}$ and $B_{i}$ represent the components of vector A and B respectively.

\subsection{Simulation Results}
We compare the proposed AESC with the traditional source and channel coding approaches over Rayleigh fading channels. We also evaluate the performance of the proposed SF mode when the source node and the destination node have different BK, which is set to different chat histories in the simulation.

Fig. \ref{model4} shows the BLEU score vs. SNR for the different schemes with the same number of transmitted symbols. For comparison, we consider a benchmark scheme, i.e., Huffman for source coding, uses Reed-Solomon (RS) (7, 5) for channel coding and binary phase shift keying (BPSK) for the modulation. We also compare with the scheme without Autoencoder. Firstly, it can be seen from the figure that the proposed AESC has a higher BLEU score than that of the scheme without Autoenoder because the Autoencoder can resist the interference of channel noise. The traditional approach performs a little better than the proposed AESC method when the SNR is above 6 dB but when SNR is less than 0 dB, the performance of the traditional approach is far less than the proposed AESC method.
Secondly, when the source node and the destination node have the same BK, the proposed AESC method with the DF mode performs better than that of the proposed AESC method with the AF mode at low SNR. When the SNR is above 0 dB, the proposed AESC method with the AF mode performs better than the proposed AESC method with the DF mode.
Thirdly, when the source node and the destination node have different BKs, the proposed SF mode has a higher BLEU score than the AF and DF modes for any SNR. The performance of the SF mode is similar to that of the DF mode when the source node and the destination node have the same BK.
\begin{figure}[t]
	\begin{center}
		\includegraphics[width=8cm]{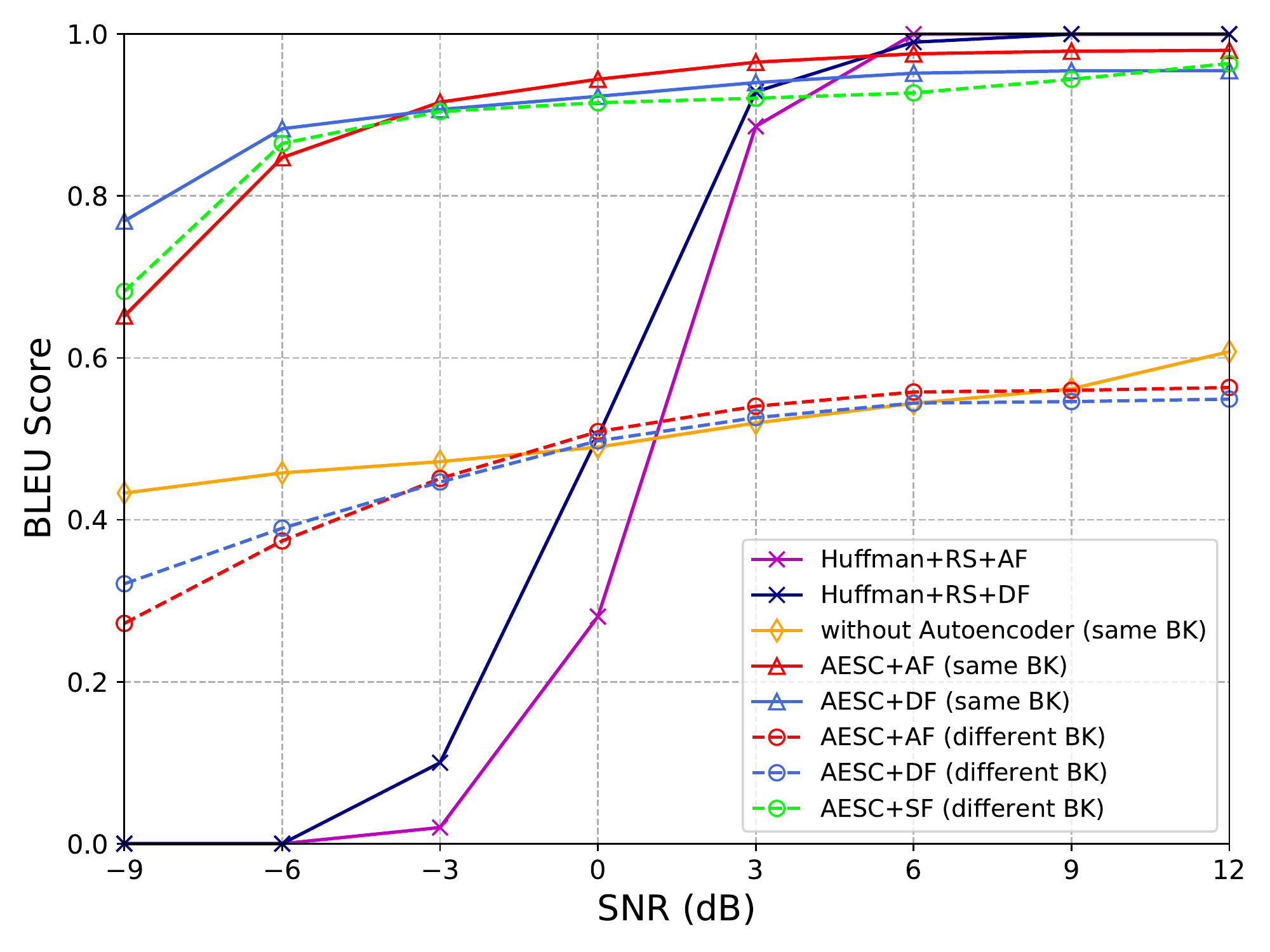}
	\end{center}
	\caption{\small{BLEU score versus SNR for the same total number of transmitted symbols in the semantic communication with relay channels for different schemes.}
	}
	\label{model4}
\end{figure}

Fig. \ref{model5} demonstrates the performances of the proposed AESC with the AF mode and DF mode in the semantic communication with relay channels when the source node and the destination node have the same BK. We fix the SNR from source to relay and change the SNR from relay to destination.
%When the SNR of both channels is higher than $5$ dB, the AF mode performs slightly better than the DF mode which the mean BLEU score is close to 1. However, when the SNR of one channel is low, especially it drops below $-5$ dB, the performance of the AF mode drops sharply.
The DF mode can maintain relatively good performance in a poor SNR region. When both SNRs are $-10$ dB, the BLEU score of the DF mode is also close to 0.8, which means that most of the semantics can be transmitted. It can also be seen that the semantic vector at the destination node with the AF mode is closer to the real transmission semantic vector in high SNR. The reason is that the distortion caused by Auto-encoder and Auto-decoder is greater than the impact of the channel in high SNR. However, when the SNR is low, the distortion of semantic vectors caused by DF mode is significantly lower than that of the AF mode.

In Fig. \ref{model6}, we evaluate the performance of different relay forward protocol via. the placement of the relay node. In the simulation, we normalize the distance between the source node and the destination node to 1, and we use $d \in(0,1)$ to denote the distance between the source node and the relay node, while $(1-\mathrm{d})$ denotes the distance between the relay node and the destination node.
 The results reveal that in the symmetrical situation of the transmit powers $P_{1}=P_{2}=5 \mathrm{~dB}$, the optimal placement of the relay node in each mode is $d = 0.5$. When $P_{1}=5 \mathrm{~dB}$ and $P_{2}=10 \mathrm{~dB}$,  the optimal relay location in each mode is $d \approx 0.4$, which means that the relay should be deployed closer to source node. We can also observe from Fig. \ref{model6} that the SF mode and the DF mode can more stably cope with the change of relay node position than the AF mode.

%Then, as mentioned in \cite{liu}, $\left|h_{1}\right|^{2}( \left|h_{2}\right|^{2})$ can be easily proved to be exponential random variable with parameter $\lambda_{1}\left(\lambda_{2}\right)$, where $\lambda_{1}=d^{4}, \text { and } \lambda_{2}=(1-d)^{4}$. We consider that $P_{1}=5 \mathrm{dBm}$ (dB in the following), $P_{2}=$ $5 \mathrm{~dB}$ or $P_{2}=$ $10 \mathrm{~dB}$ with self-interferences are set to $\sigma^{2}=1$.

%This is because the decoding error caused by the previous channel noise will be amplified, which is worse than the decoding error caused by the latter channel. So when using DF mode, the first channel condition should not be too bad.

\begin{figure}[t]
	\begin{center}
		\includegraphics[width=8.8cm]{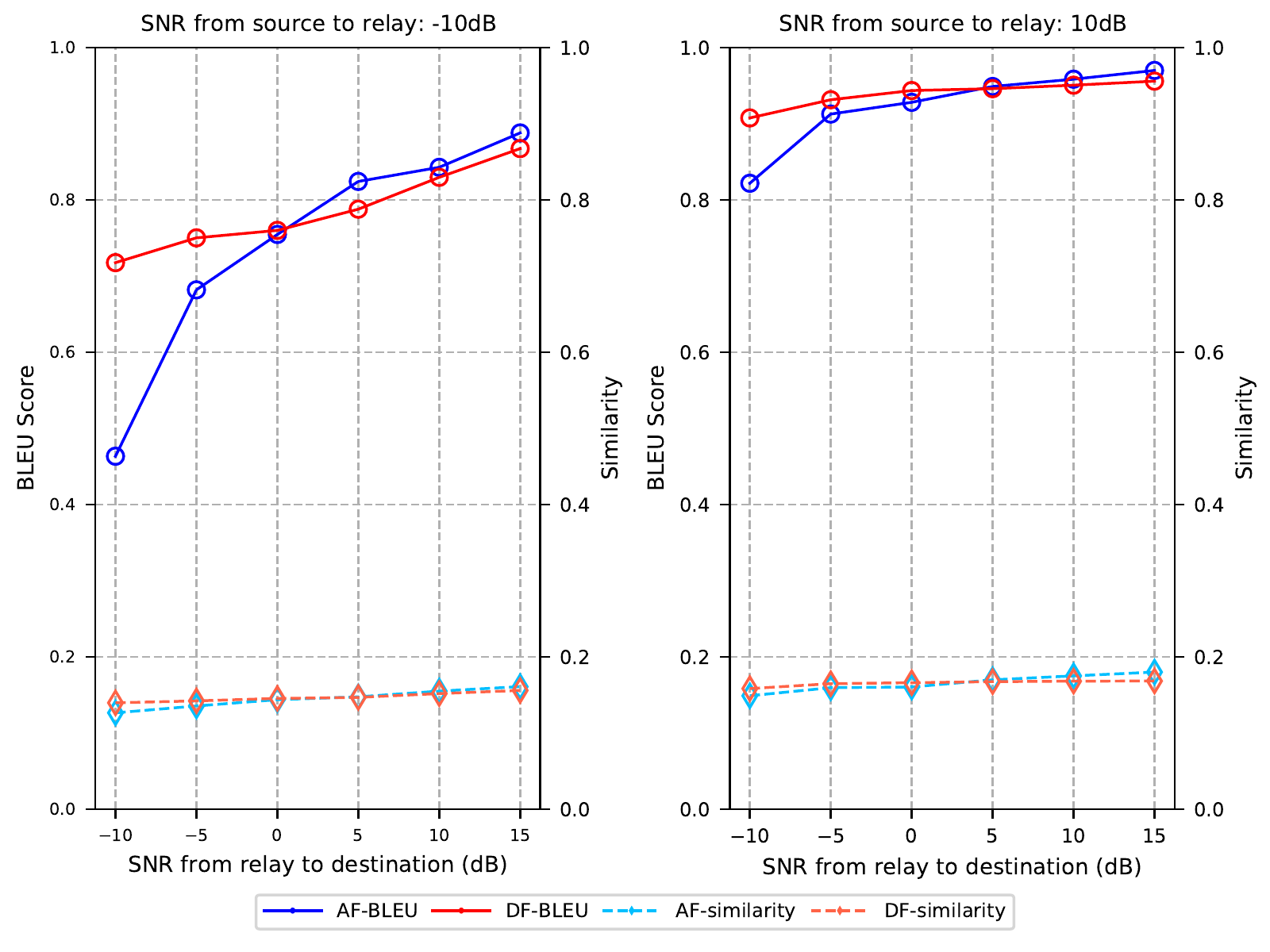}
	\end{center}
	\caption{\small{BLEU score and sentence similarity versus SNR from relay to destination for the proposed AESC with the AF mode and the DF mode.}
	}
	\label{model5}
\end{figure}
\begin{figure}[t]
	\begin{center}
		\includegraphics[width=8cm]{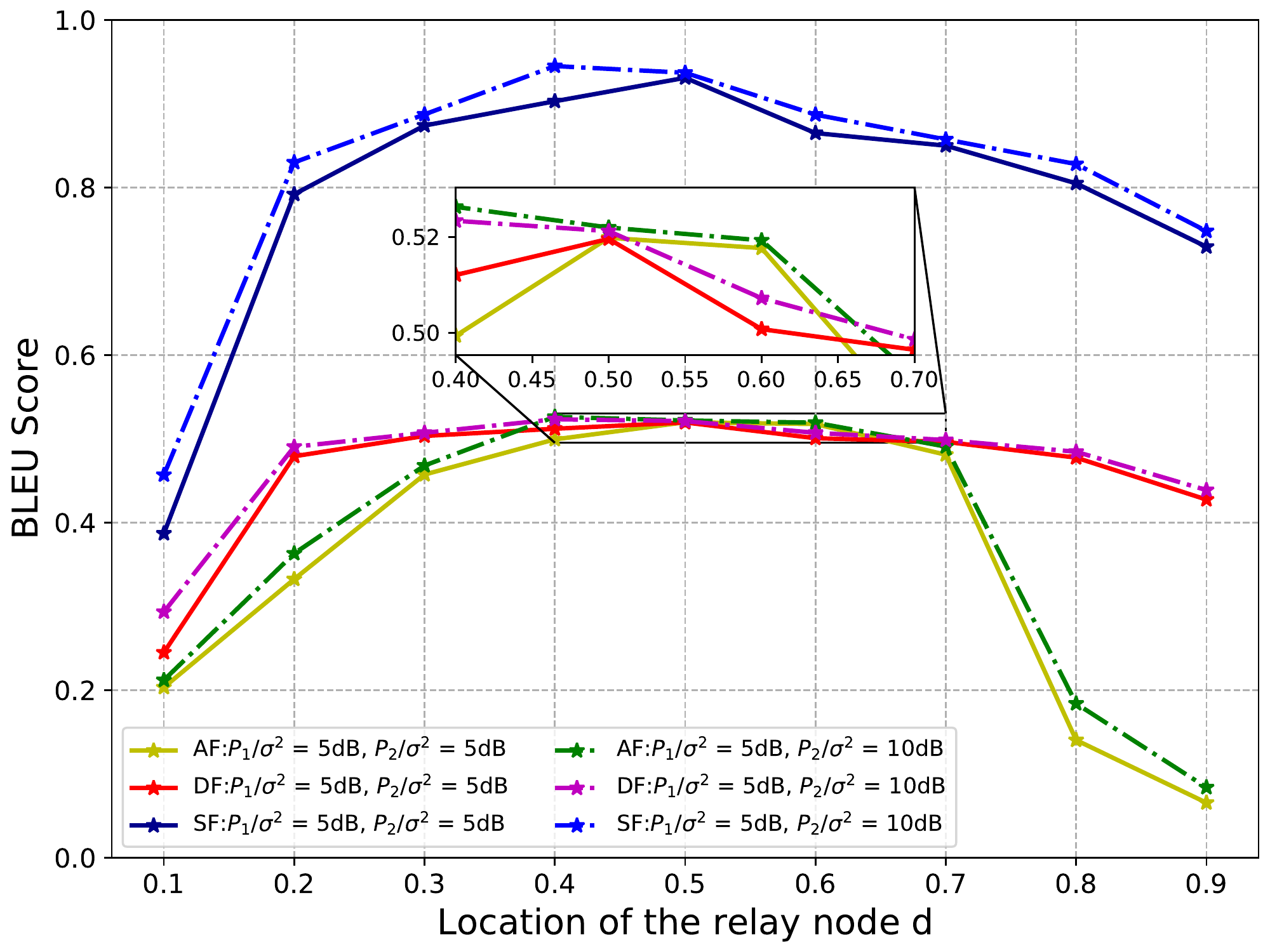}
	\end{center}
	\caption{\small{BLEU score versus the relay location $d$ in the AF mode, the DF mode, and the SF mode. Here, we use $\sigma^{2}=1$.}
	}
	\label{model6}
\end{figure}

\section{Conclusion}
This letter proposed a semantic communication approach based on Autoencoder for the wireless relay channel with a novel semantic forward protocol. The proposed Autoencoder-based approach uses Autoencoder to extract and compress semantic information and reconstructs its semantic features. For the proposed SF protocol, the relay node can cooperatively use the background knowledge of the source node and the destination node to forward semantic information at the semantic level, to solve the problem of direct semantic communication under different background knowledge between the source node and the destination node. Simulation results demonstrated the effectiveness and advancement of the proposed approach.

%The proposed scheme used the attention mechanism to encode text sentences in semantic dimension, and the Autoencoder can extract and compress semantic information and reconstruct its semantic features.
%We demonstrated through experiments that the Autoencoder greatly improved the anti-noise performance of the AESC. Moreover, we evaluated the performance of the proposed AESC with the AF mode and the DF mode in the wireless relay channel when sender and receiver have the same BK, and we verified that the SF mode greatly reduces the semantic distortion compared to AF and DF when sender and receiver have different BK. The valuable insights on the impact of the key system parameters on the proposed AESC with relay channels are obtained for future application.

\bibliographystyle{IEEEtran}
\bibliography{ref}{}
\end{document}